\documentclass[runningheads]{llncs}

\usepackage{graphicx}
\usepackage{listings}
\usepackage{color}
\definecolor{javared}{rgb}{0.6,0,0} 
\definecolor{javagreen}{rgb}{0.25,0.5,0.35} 
\definecolor{javapurple}{rgb}{0.5,0,0.35} 
\definecolor{javadocblue}{rgb}{0.25,0.35,0.75} 
 
\usepackage{tikz}
\usepackage{rotating}
\usepackage{inconsolata}
\usepackage{amsfonts}
\usepackage{pifont}
\newcommand{\xmark}{\ding{55}}
\newcommand{\cmark}{\ding{51}}
\begin{document}

\lstset{
  basicstyle=\scriptsize\ttfamily
}

\title{A Systematic Study on Static Control Flow Obfuscation Techniques in Java}

\author{Renuka Kumar, Anjana Mariam Kurian}
\institute{Amrita Center for Cybersecurity Systems \& Networks, Amritapuri Campus,\\ 
Amrita Vishwa Vidyapeetham}
\maketitle

\begin{abstract}
Control flow obfuscation (CFO) alters the control flow path of a program without altering its semantics. Existing literature has proposed several techniques; however, a quick survey reveals a lack of clarity in the types of techniques proposed, and how many are unique. What is also unclear is whether there is a disparity in the theory and practice of CFO. In this paper, we systematically study CFO techniques proposed for Java programs, both from papers and commercially available tools. We evaluate 13 obfuscators using a dataset of 16 programs with varying software characteristics, and different obfuscator parameters. Each program is carefully reverse engineered to study the effect of obfuscation. Our study reveals that there are 36 unique techniques proposed in the literature and 7 from tools. Three of the most popular commercial obfuscators implement only 13 of the 36 techniques in literature. Thus there appears to be a gap between the theory and practice of CFO. We propose a novel classification of the obfuscation techniques based on the underlying component of a program that is transformed. We identify the techniques that are potent against reverse engineering attacks, both from the perspective of a human analyst and an automated program decompiler. Our analysis reveals that majority of the tools do not implement these techniques, thus defeating the protection obfuscation offers. We furnish examples of select techniques and discuss our findings. To the best of our knowledge, we are the first to assemble such a research. This study will be useful to software designers to decide upon the best techniques to use based upon their needs, for researchers to understand the state-of-the-art and for commercial obfuscator developers to develop new techniques.


\end{abstract}

\section{Introduction}\label{sec:Introduction}
Obfuscation is a program transformation technique that renders a program unreadable to a human while preserving its semantics. A software developer uses static code obfuscation techniques to secure code and data from reverse engineering attacks. There are three types of obfuscation- lexical, data and control flow obfuscation. Lexical obfuscation replaces class, field and method identifiers with new random identifiers or words from a dictionary. Data obfuscation modifies data structures used by a program. Control flow obfuscation (CFO) alters the control flow path of a program. Malware authors use obfuscation to impede security analysts from reverse engineering their code or to evade automated program analyzers. 

In this paper, we focus on control flow obfuscation techniques. While existing literature proposes several techniques, the differences between them is unclear. What is also unknown is if there is a disparity in the theory and practice of CFO. This knowledge can aid software developers to decide on the best techniques to use to secure their code. Commercial obfuscator developers and researchers can use this knowledge to advance the state-of-the-art and steer research in this area.




There are a few existing surveys on control flow obfuscation \cite{RefJ26},\cite{RefJ19},\cite{RefJ27}, \cite{RefJ24},\cite{RefJ11} \cite{RefJ16}. Colberg et al. propose groundbreaking work on the taxonomy of obfuscating transformations. They classify 15 control flow transformations into four types - Opaque Predicate, Computation Transformation, Aggregation Transformation and Ordering Transformations. However, their classification is based on theoretical foundations alone. Schrittwieser et al. \cite{RefJ16} classifies obfuscation techniques into three categories- data obfuscation, static code rewriting and dynamic code rewriting. They survey literature to analyze the strength of obfuscation against common adverserial goals. 

Balakrishnan et al. \cite{RefJ24} presents a literature survey of all code obfuscation techniques used to thwart static analysis and program disassembly. However, all the CFO techniques they propose is also discussed by Colberg et al. \cite{RefJ26}. You et al. \cite{RefJ11} discusses a survey of obfuscation techniques in Windows malware binaries. Majumdar et al. \cite{RefJ27} survey the strengths and weaknesses of two CFO techniques - control flow flattening and opaque predicates. Xu et al. \cite{RefJ57} compares and contrasts code-oriented and model-oriented obfuscation techniques to conclude that there are no secure and usable obfuscation techniques. Lim \cite{RefJ54} evaluates various code obfuscation techniques and measures performance of each. Low et al. \cite{RefJ19} discusses ways to create resilient and stealthy opaque predicates. 


In this paper, we systematize knowledge of control flow obfuscation on Java programs by studying the techniques proposed in the literature and those implemented by obfuscation tools. We assemble a list of known techniques into a single study and classify it into 5 categories based on the component (or level) of a program on which the transformation is applied. Though we use the taxonomy proposed by Colberg et al. \cite{RefJ26} as the basis of our research, we deviate substantially in our classification. We evaluate 13 obfuscation tools and determine that only 6 are viable for analysis. The remaining obfuscators were either unavailable for download, end-of-lived or had unresolvable program errors that made it unusable. Obfuscators used for Android programs can also be used for Java applications. However, certain Android obfuscators will not work on Java bytecode. Hence we evaluate tools that can obfuscate Java programs only. 

In order to evaluate obfuscators, we choose a sample set of 16 programs with varying control flow properties. Each obfuscator is repeatedly applied to the 16 sample programs by varying their configuration parameters. The bytecode of the obfuscated program is then manually analyzed and compared with that of the original program to tabulate our findings. Wherever possible, we confirm our results with any documentation provided by the tool. We apply no lexical or data obfuscation and restrict our analysis to CFO only. 

From our study, we identified a total of 43 unique techniques, 36 from literature and 7 from tools. Of the 36 techniques from literature, 12 are not implemented in any of the tools. Three of the most popular commercial obfuscators implement only 13 of the 36 techniques from the literature. Thus there appears to be a gap between the theory and practice of CFO. We assess the potency of each technique from the perspective of a human analyst and a program decompiler. We determine that techniques that are applied on a basic block are the most potent. However, majority of the techniques do not implement them, thus defeating the reliability of obfuscation. We identify the most popular techniques in tools and identify those that are incorrectly cited as control flow obfuscating. In the end, we discuss our findings from this study.

To summarize, we make the following contributions:
\begin{itemize}
    \item We survey literature for CFO techniques and identify 36 unique techniques from literature. We furnish examples of select techniques and identify those that are incorrectly cited as control flow obfuscating.
    \item We evaluate 13 obfuscators and identify the 6 viable tools that can control flow obfuscate Java programs. 
    \item We evaluate each tool on a set of 16 programs with varying software characteristics by repeated application of each obfuscator with varying configuration parameters. We identify 7 novel techniques. Three of the most popular commercial obfuscators implement only 13 of the 36 techniques. Thus there appears to be a gap between the theory and practice of CFO.  We also identify the most commonly applied obfuscation techniques in tools. 
    \item We propose a classification of the 43 techniques based on the component of a program that is transformed. We discuss the potency of the techniques, and identify those that are resilient to decompilers. We determine that majority of the obfuscators do not implement these techniques, thus defeating its purpose.
\end{itemize}

\section{Running Example}\label{sec:Runningeg}
Listing \ref{code:running} is an iterative implementation of a binary search program in Java. The program will be used as the running example for the remainder of this paper. The example has all the components present in a typical program such as \textit{if-else} condition, loop etc. The function \textit{binarySearch} takes two parameters as input - an array and the element to be searched. It returns the index at which the element was found and $-1$ otherwise. 

\section{Techniques in Literature}\label{sec:TechfrmLit}
In this section we discuss techniques found in the literature grouped by the obfuscating paradigm used. Wherever possible, we provide examples of each technique using the running example. The code snippet on the left shows the original code. Obfuscated code is on the right. Examples for some techniques have been omitted when it is similar to already stated techniques or when there are other literature that cites them.

\begin{lstlisting}[caption=Running Example, label={code:running}] code
class FindElement{
   static int binarySearch(int arr[], int x){
       int l = 0, r = arr.length - 1;
       while (l <= r){
           int m = l + (r-l)/2;
           if (arr[m] == x)  // Check if x is present at mid
               return m+1;
           if (arr[m] < x) // If x greater, ignore left half
               l = m + 1;
           else // If x is smaller, ignore right half
               r = m - 1;
       }
       return -1;
   }
}
class BinarySearch extends FindElement{
    public void printResult(int result,boolean isPresent) {
       if(isPresent==false)
           System.out.println("Element not present");
       else
           System.out.println("Element was found at index " + result);
   }
   public static void main(String args[]){
       FindElement ob1=new FindElement();
       BinarySearch ob2=new BinarySearch();
       boolean isPresent=false;
       int arr[] = new int[args.length];
       if (args == null) return;
       for (int i = 0; i < args.length; i++) 
         arr[i] = Integer.parseInt(args[i]);

       //print the array
       for (int i =0; i < arr.length; i++)
           System.out.println(arr[i]);
       int index = ob1.binarySearch(arr, 10);
       if(index!=-1){
           isPresent=true;
           ob2.printResult(index,isPresent);
       }
       else
           ob2.printResult(index,isPresent);
   }
}
\end{lstlisting}

\subsection{Using Opaque Predicates}
An opaque predicate is a predicate whose value is known apriori to the obfuscator. There are several ways in which an opaque predicate can be used to obfuscate a program \cite{RefJ1}, \cite{RefJ2}, \cite{RefJ3}, \cite{RefJ4}, \cite{RefJ5}, \cite{RefJ10}, \cite{RefJ12}, \cite{RefJ13}, \cite{RefJ16}, \cite{RefJ17}, \cite{RefJ21}, \cite{RefJ26}, \cite{RefJ27}, \cite{RefJ28}, \cite{RefJ29}, \cite{RefJ30}, \cite{RefJ40}, \cite{RefJ41}, \cite{RefJ42}, \cite{RefJ43}, \cite{RefJ47}, \cite{RefJ48}, \cite{RefJ49}, \cite{RefJ50}, \cite{RefJ54}. Below we discuss the techniques that use opaque predicates. Existing literature also proposes algebraic techniques to create resilient opaque predicates; however, we do not study resilience of predicates in this work. 

\begin{enumerate}
    \item \textit{Extending conditionals.} An opaque predicate is inserted into a conditional expression of either an \textit{if-else} statement or a loop. Here inserting the opaque predicate does not alter the evaluation of the expression \cite{RefJ25}, \cite{RefJ26}. 
    \item \textit{Adding Redundant Operands.} Opaque variables or expressions are added as operands to arithmetic expressions \cite{RefJ25}, \cite{RefJ26}.  
    \item \textit{Dead Code Insertion.} An opaque predicate that guards a set of statements is inserted into the program in such a way that the code block guarded by the predicate is never executed (dead code) \cite{RefJ7}, \cite{RefJ10}, \cite{RefJ11}, \cite{RefJ15}, \cite{RefJ16}, \cite{RefJ20}, \cite{RefJ26}, \cite{RefJ46}, \cite{RefJ47}, \cite{RefJ50}.  
\end{enumerate}

The example below shows all three techniques discussed above. In the obfuscated program, the original \textit{if} statement is modified by adding an opaquely \textit{false} predicate denoted by the variable \textit{bool}. The arithmetic expression of the \textit{return} statement is changed to also use a redundant operand called \textit{redop}. The obfuscated code also contains an additional \textit{if} statement guarded by the predicate \textit{bool} whose consequent never gets executed.

\noindent\begin{minipage}{.48\textwidth}
\begin{lstlisting}[frame=tlrb,label=code:example-2-1, numbers=none]{Deadcode}
if (arr[m] == x)
    return m+1;
\end{lstlisting}
\end{minipage}\hfill
\begin{minipage}{.48\textwidth}
\begin{lstlisting}[frame=tlrb,label=code:example-2-2,numbers=none]{Deadcode}
boolean bool;
int redop = 1;
if (arr[m] == x && !bool)
    if(bool){ 
        /*deadcode*/
        return 0;
    } 
    return (m+1) * redop;
}
\end{lstlisting}
\end{minipage}

\subsection{Irrelevant Code Insertion}
In this technique, the obfuscator inserts a set of irrelevant statements into the program \cite{RefJ16}. Unlike \textit{Dead Code Insertion}, the program executes the inserted statements. However, they may be irrelevant to it, or not present in the original program. The string variable \textit{str} is the irrelevant code in the example.

\noindent\begin{minipage}{.48\textwidth}
\begin{lstlisting}[frame=tlrb,label=code:example-3-1, numbers=none]{Exprreo}
if (isPresent == false)
   System.out.println("Element 
                 not present");
\end{lstlisting}
\end{minipage}\hfill
\begin{minipage}{.48\textwidth}
\begin{lstlisting}[frame=tlrb,label=code:example-3-2,numbers=none]{Exprreo}
if (isPresent == false) {
   String str="Hello World!";
   System.out.println("Element 
                 not present"); 
}
\end{lstlisting}
\end{minipage}

\subsection{Ordering obfuscation.}
Ordering obfuscation changes the order of evaluation of expressions, loops, statements or methods while preserving dependencies \cite{RefJ10}, \cite{RefJ16}, \cite{RefJ18}, \cite{RefJ25}, \cite{RefJ26}, \cite{RefJ54}.

\begin{enumerate}
    \item \textit{Reordering Expressions.} This technique changes the order of evaluation of sub-expressions without changing its valuation. The example below shows how to obfuscate a conditional of the \textit{if} statement.
    
    \begin{minipage}{.46\textwidth}
\begin{lstlisting}[frame=tlrb,label=code:example-4-1,numbers=none]{Exprreo}
if (arr[m] < x) 
    l = m + 1;
\end{lstlisting}
\end{minipage}\hfill
\begin{minipage}{.46\textwidth}
\begin{lstlisting}[frame=tlrb,label=code:example-4-2,numbers=none]{Exprreo}
if (x >= arr[m]) 
    l = m + 1;
\end{lstlisting}
\end{minipage}

    \item \textit{Reordering Loop.} The example shows how to change the evaluation of a loop index.

\noindent\begin{minipage}{.46\textwidth}
\begin{lstlisting}[frame=tlrb,label=code:example-5-1,numbers=none]{Loopreo}
for(int i=0;i<arr.length;i++) 
    System.out.println(arr[i]);
\end{lstlisting}
\end{minipage}\hfill
\begin{minipage}{.46\textwidth}
\begin{lstlisting}[frame=tlrb,label=code:example-5-2,numbers=none]{Loopreo}
for(int i=arr.length-1;i>=0;i--) 
    System.out.println(arr[i]);
\end{lstlisting}
\end{minipage}

    \item \textit{Reordering Statements.} In this technique, an obfuscator reorders statements within a basic block to preserve dependencies \cite{RefJ46}, \cite{RefJ50}.

\item \textit{Reordering Code Blocks.} This technique reorders basic blocks while preserving dependencies.

\item \textit{Reordering Methods.} This techniques changes the order of its subroutines \cite{RefJ11}, \cite{RefJ46}. This technique can generate n! different variants of a program, where n is the number of subroutines. This approach moves the structural position of the method in a file.


\end{enumerate}

\subsection{Instruction Substitution.}
In this technique, the obfuscator replaces a sequence of instructions with an alternate sequence of semantically equivalent instructions \cite{RefJ7}, \cite{RefJ11}, \cite{RefJ20}, \cite{RefJ46}. For instance, the \textit{goto} at line 47 is replaced with \textit{iload\_2} and \textit{ifeq} instructions. 

\noindent\begin{minipage}{.46\textwidth}
\begin{lstlisting}[frame=tlrb,label=code:example-7-1,numbers=none]{Instrsub}
47 : goto 107
50 : iload_1
107: return
\end{lstlisting}
\end{minipage}\hfill
\begin{minipage}{.50\textwidth}
\begin{lstlisting}[frame=tlrb,label=code:example-7-2,numbers=none]{Instrsub}
66  : iload_2
67  : ifeq 129
70  : iload_1
129 : return
\end{lstlisting}
\end{minipage}

A specific implementation of this technique has been mentioned in \cite{RefJ59} called the \textit{Replacing Goto} obfuscation. In this technique \textit{goto} instructions are replaced with conditional branch instructions. 

\subsection{Control Flow Flattening}
This technique works by first flattening the control flow graph of a program such that all the basic blocks have the same successor and predecessor. A dispatcher variable is then introduced to guide the control flow of the program \cite{RefJ2}, \cite{RefJ4}, \cite{RefJ5}, \cite{RefJ7}, \cite{RefJ100}, \cite{RefJ12}, \cite{RefJ13}, \cite{RefJ14}, \cite{RefJ21}, \cite{RefJ22}, \cite{RefJ25}, \cite{RefJ40}, \cite{RefJ41}, \cite{RefJ43}, \cite{RefJ44}, \cite{RefJ45}, \cite{RefJ56}. In the example, the original code snippet has two basic blocks. The control flow flattened program encapsulates both the basic blocks into a \textit{switch} block with each basic block forming a \textit{case} statement. The dispatcher variable here is \textit{var}. When \textit{var = 2}, control flow enters the second case statement that contains the first basic block.

\noindent\begin{minipage}{.48\textwidth}
\begin{lstlisting}[frame=tlrb,label=code:example-8-1, numbers=none]{CFF}
int arr[] = {2, 3, 4, 10, 40};
for (int i =0; i < arr.length; i++) 
    System.out.println(arr[i]);
\end{lstlisting}
\end{minipage}\hfill
\begin{minipage}{.48\textwidth}
\begin{lstlisting}[frame=tlrb,label=code:example-8-2,numbers=none]{CFF}
int var=2;
while(var!=0){ 
 switch(var){ 
  case 1:
     for(int i =0;i<arr.length;i++) 
       System.out.println(arr[i]);
     var=0;
     break;
  case 2:
     int arr[] = {2, 3, 4, 10, 40};
     var=1;
     break;
  default:break; 
  } 
}
\end{lstlisting}
\end{minipage}

\subsection{Method Transformation}
This technique obscures method invocations. There are four types of transformation: 

\begin{enumerate}
\item \textit{Inline Methods.} The technique replaces a method invocation with the body of the method \cite{RefJ1}, \cite{RefJ6}, \cite{RefJ10}, \cite{RefJ15}, \cite{RefJ18}, \cite{RefJ24}, \cite{RefJ26}, \cite{RefJ50}, \cite{RefJ54}. For instance, the body of the function \textit{printResult} replaces its invocation (line 17) in the running example. 

\item \textit{Outline Methods.} This technique is the inverse of inlining; it replaces a sequence of statements with a function call \cite{RefJ1}, \cite{RefJ6}, \cite{RefJ10}, \cite{RefJ24}, \cite{RefJ26}, \cite{RefJ50}, \cite{RefJ54}.

\item \textit{Clone Methods.} This obfuscation modifies a method’s call site by invoking clones of the same method \cite{RefJ1}, \cite{RefJ16}, \cite{RefJ24}, \cite{RefJ26}. The call to \textit{binarySearch} is obfuscated by inserting clones of the same function.

\noindent\begin{minipage}{.43\textwidth}
\begin{lstlisting}[frame=tlrb,label=code:example-9-1,numbers=none]{Clone}
int result = binarySearch(arr, 10);
\end{lstlisting}
\end{minipage}\hfill
\begin{minipage}{.48\textwidth}
\begin{lstlisting}[frame=tlrb,label=code:example-9-2,numbers=none]{Clone}
int choice=1;
int result;
if(choice==1)
  result = binarySearch1(arr, 10);
else
  result = binarySearch2(arr, 10);
\end{lstlisting}
\end{minipage}

\item \textit{Interleave Methods.} The technique merges two separate methods into a single method \cite{RefJ1},  \cite{RefJ26}. 
\end{enumerate}
\subsection{Replacing if(non) null instructions with try-catch blocks}
The instructions \textit{ifnull} and \textit{ifnonnull} are used for obfuscation. The \textit{ifnull} instruction checks if the topmost element on the stack is null, whereas the \textit{ifnonnull} instruction checks if the top most instruction is non-null. These instructions then jump to a location as specified by a label. These instructions are replaced by \textit{try-catch} constructs to alter the control flow. This may be done either by inserting a statement that will cause an exception or by explicitly throwing an exception \cite{RefJ37}. Below is an example of the former.

\noindent\begin{minipage}{.48\textwidth}
\begin{lstlisting}[frame=tlrb,label=code:example-10-1, numbers=none]{Trycatch}
if (args == null) return;
for (int i = 0; i < args.length; i++) 
  arr[i] = Integer.parseInt(args[i]);

\end{lstlisting}
\end{minipage}\hfill
\begin{minipage}{.48\textwidth}
\begin{lstlisting}[frame=tlrb,label=code:example-10-2,numbers=none]{Trycatch}
try {
  for (int i = 0; i < args.length; i++) 
    arr[i] = Integer.parseInt(args[i]);
} catch (NullPointerException ne) {
  return;
}
\end{lstlisting}
\end{minipage}

\subsection{Loop Transformations}
These transformations manipulate loops without changing their behaviour \cite{RefJ16}, \cite{RefJ26}. It is of four types.

\begin{enumerate}
    \item \textit{Loop Fission.} The technique splits a loop into multiple loops \cite{RefJ21}. For instance, the \textit{for} loop in the original program is split into two by causing the loops to iterate only half the array.

\begin{minipage}{.46\textwidth}
\begin{lstlisting}[frame=tlrb,label=code:example-11-1,numbers=none]{Loopfis}
for(int i=1;i<arr.length;i++)   
    System.out.println(arr[i]);
\end{lstlisting}
\end{minipage}\hfill
\begin{minipage}{.46\textwidth}
\begin{lstlisting}[frame=tlrb,label=code:example-11-2,numbers=none]{Loopfis}
for(int i=1;i< arr.length/2;i++) 
    System.out.println(arr[i]);
for(int i=arr.length/2;
    i< arr.length;i++) 
    System.out.println(arr[i]);
\end{lstlisting}
\end{minipage}

\item \textit{Loop Blocking.} This is a well-known optimization technique used by compilers for caching. Here, a loop is partitioned into small blocks. For instance, the loop in the original program is blocked by a block of size 2.

\begin{minipage}{.46\textwidth}
\begin{lstlisting}[frame=tlrb,label=code:example-12-1,numbers=none]{Loopbloc}
for(int i=0;i< arr.length;i++) 
    System.out.println(arr[i]);
\end{lstlisting}
\end{minipage}\hfill
\begin{minipage}{.46\textwidth}
\begin{lstlisting}[frame=tlrb,label=code:example-12-2,numbers=none]{Loopbloc}
for (int j=0;j<arr.length;j+=2)
 for(int i=j; j< min(
    arr.length, j+2); i++)
     System.out.println(arr[i]);
\end{lstlisting}
\end{minipage}

\item \textit{Loop Unrolling.} Also known as \textit{loop unwinding}, this is yet another compiler optimization that trades space for execution speed\cite{RefJ6}, \cite{RefJ10}, \cite{RefJ54}. It reduces the number of iterations of a loop by increasing the size of the program.  For instance, in the obfuscated program, the \textit{print} statement is unrolled.

\noindent\begin{minipage}{.46\textwidth}
\begin{lstlisting}[frame=tlrb,label=code:example-13-1,numbers=none]{LoopUnrol}
for (int i =0; i < arr.length; i++) 
   System.out.println(arr[i]);
\end{lstlisting}
\end{minipage}\hfill
\begin{minipage}{.46\textwidth}
\begin{lstlisting}[frame=tlrb, label=code:example-13-2,numbers=none]{LoopUnrol}
for (int i=0;i<arr.length; i=i+3) {
    System.out.print(arr[i]);
    System.out.println(arr[i+1]);
    System.out.println(arr[i+2]);
}
\end{lstlisting}
\end{minipage}

\item \textit{Replace with Equivalent Codes.} This obfuscation inserts, modifies or removes statements, identifiers, and literals \cite{RefJ38}. This is also called as Code Clone III. In the example below, the elemet \textit{arr[i]} is assigned to another variable \textit{num}.

\begin{minipage}{.46\textwidth}
\begin{lstlisting}[frame=tlrb,label=code:example-23-1,numbers=none]{Clone3}
for (int i =0; i < arr.length; i++) {
   System.out.println(arr[i]);
}
\end{lstlisting}
\end{minipage}\hfill
\begin{minipage}{.46\textwidth}
\begin{lstlisting}[frame=tlrb, label=code:example-23-2,numbers=none]{Clone3}
for (int i =0; i < arr.length; i++){
   int num = arr[i];
   System.out.println(num);
}
\end{lstlisting}
\end{minipage}

\item \textit{Code Clone-Type IV.} Here code snippets are modified to produce syntactically different, yet semantically identical variants \cite{RefJ38}. For instance, an obfuscator replaces a  \textit{for} loop with a \textit{while} loop.

\item \textit{Inserting Dummy Loop.} In this technique empty loops are inserted into the code. An alternative implementation is combining basic blocks into a loop.

\item \textit{Intersecting Loops.}This approach inserts intersecting loops into the control flow and guarded by an opaque predicate to skip the loop \cite{RefJ59}. The opaque predicate causes the the inserted code to act like dead code. 

\end{enumerate}

\subsection{Basic Block Fission}
This technique splits the chosen code blocks into finer pieces and inserts opaque predicates and \textit{goto} instructions in them \cite{RefJ59}. The use of \textit{goto} may cause the decompiler to fail in this case. 

\subsection{Remove Library and Program Idioms}
The technique uses obfuscated version of library functions and removes/replaces programming idioms from the original program \cite{RefJ16}, \cite{RefJ25}, \cite{RefJ26}. For example, \textit{System.out.println} library call is replaced with a semantically equivalent implementation that use \textit{BufferedWriter}.

\noindent\begin{minipage}{.48\textwidth}
\begin{lstlisting}[frame=tlrb,label=code:example-16-1,numbers=none]{Remlib}
System.out.println("Element not 
present");
\end{lstlisting}
\end{minipage}\hfill
\begin{minipage}{.48\textwidth}
\begin{lstlisting}[frame=tlrb,label=code:example-16-2,numbers=none]{Remlib}
BufferedWriter bw = new 
BufferedWriter(new 
OutputStreamWriter(System.out));
bw.write("Element not present");
\end{lstlisting}
\end{minipage}

\subsection{Convert a Reducible to Non-Reducible Flow-graph}
In this technique, instructions that have no direct correspondence with the source language is used to obfuscate a program. This makes the transformation language-breaking or non-reducible \cite{RefJ26}. A commonly used approach is to use the \textit{goto} statement to express arbitrary control flow. This makes it a non-reducible flow-graph as the Java language itself can only express structured control flow (reducible flow-graph). This can easily be done by inserting a bogus jump to the middle of the body of a \textit{while} loop using an opaque predicate that will never get executed. When the loop has multiple headers or entry points, it no longer becomes structured and hence will become non-reducible. 

\subsection{Table Interpretation}
This technique is also called as \textit{Virtualization} \cite{RefJ8}, \cite{RefJ9}, \cite{RefJ23}, \cite{RefJ48}, \cite{RefJ26}. A program's bytecode is converted into a custom instruction set that is executed by a VM interpreter included within the obfuscated application. This transformation induces additional runtime over head and hence is reserved only for extremely sensitive portions of an application. 
\subsection{Parallelizing the Code}
\label{sec:11}
In this technique, multiple processes are created to execute code blocks that have no data dependencies. This may be done either by creating dummy processes and running them in parallel with the instructions of the program or by splitting a sequential section of the application into multiple sections that can be executed in parallel. Code block that have data dependencies can be parallelized using synchronizing functions \cite{RefJ15}, \cite{RefJ16}, \cite{RefJ25}, \cite{RefJ26}.

\renewcommand{\arraystretch}{1.2}
\begin{table}
\caption{List of Control Flow Obfuscators in the Study}
\label{tab:listoftools}
\scriptsize
\begin{tabular}{p{2.7cm}p{1.6cm}p{2.2cm}p{2.5cm}p{1.5cm}}
\hline\noalign{\smallskip}
Obfuscator & Software Version & Availability & Input & Evaluated \\
\noalign{\smallskip}\hline\noalign{\smallskip}
Zelix Klassmaster & 6.1 & Paid version & Bytecode & Y \\
Allatori & 6.4 & Trial version & Bytecode & Y  \\
DashO  & 8.3.0 & Trial version & Bytecode & Y  \\
JBCO & 2.5.0 & Not-Available & Bytecode & N \\
JMOT & 3.0 & Open-source & Bytecode & Y \\
Sandmark & 2.1 & Open-source & Bytecode & Y \\
Jfuscator & Unspecified & Trial version & Byte/Sourcecode & Y\\ 
JOAD & 2.1 & Open-source & Bytecode & N\\
JShield & Unspecified & Not available & Source code & N\\            
Arxan & Unspecified & Paid version & Bytecode & N\\                    
Cloakware & Unspecified & Not available & Sourcecode & N\\          
Smokescreen & 3.11 & Not available & Bytecode & N\\                 
Codeshield & 2.0 & Not available & Bytecode & N\\
\noalign{\smallskip}\hline
\end{tabular}
\end{table}

\section{Tools Evaluation}\label{sec:ToolEval}
We identify 13 tools as listed in table \ref{tab:listoftools} that support control flow obfuscation. We restrict our evaluation to control flow obfuscators and exclude lexical and data obfuscators such as Proguard. Of the 13 tools Only 6 tools are viable - DashO, Allatori, Zelix Klassmaster, Sandmark, JMOT, and JFuscator. JBCO is the only tool which we could not evaluate, yet is included since they furnish a detailed documentation of their techniques. For those that were not evaluated, we indicate why. In case of paid proprietary tools, we use their trial versions for evaluation, except for Zelix Klassmaster, for which we acquired a licensed version. 

In order to conduct the study, we use a set of 16 synthetic programs with and without control flow constructs. We intended the programs to contain a heterogeneous mix of properties. The programs have one or more of the following characteristics: 1) non-nested \textit{if} statements 2) non-nested \textit{if-else} statements 3) one/many \textit{for} or \textit{while} loops 4) \textit{switch-case} statements 5) nested conditional statements 6) nested \textit{try-catch} blocks 7) use of complex data structures such as \textit{hashmaps} 8) variable number of methods 9) accepts user inputs. In a case where the conditionals are nested, we introduce a nesting level of up to 3 levels.

The source/bytecode of each of the 16 programs are repeatedly supplied as input to each control flow obfuscator for different obfuscator parameters. We manually manually examine the bytecode output to identify the differences between the original and the transformed programs using the \textit{javap} command. This is because, the decompiled output of a program may vary depending on the decompiler used. Certain obfuscations also produce programs that cannot be decompiled.  Lastly, while some of the techniques do not appear to alter the control flow of the program from the perspective of source code, it is still a control flow obfuscation if it alters the control flow of the bytecode. We use \textit{JD-GUI} decompiler wherever possible to verify our findings.  

Table \ref{tab:techniquesintools} details the list of obfuscation techniques supported by each tool. 

\subsection{Zelix Klassmaster}
Zelix supports 3 levels of control flow obfuscation- light, normal and aggressive, and data obfuscation. Zelix uses only techniques proposed by the literature. They using static boolean variables as opaque predicates and insert emtpty \textit{if} statements guarded by opaquely false predicates. Zelix also reorders the constant pool index, which is a structural obfuscation technique.


\subsection{Allatori}
This obfuscator is available as a plugin for Eclipse. It supports incremental obfuscation of code to allow for consistency with the previous obfuscated versions of a program. 

\iftrue
\renewcommand{\arraystretch}{1.3}
\begin{table}[ht!]
\caption{List of Techniques from Tools}
\label{tab:techniquesintools}
\scriptsize
\noindent\begin{minipage}[b]{0.30\linewidth}
\begin{tabular}{p{1.5cm}p{4.2cm}}
\hline\noalign{\smallskip}
Obfuscator & Obfuscation Technique  \\
\noalign{\smallskip}\hline\noalign{\smallskip}
Zelix  &  Irrelevant Code Insertion\\
                  &  Extending Conditionals\\
                  &  Opaque Branch Insertion \\
                  & Opaque Predicate Insertion \\
                  &  Dead Code Insertion\\
                  &  Instruction Substitution \\
                  &  Convert Reducible to Irreducible Flowgraph \\
                  &  Replacing if(non) Null Instructions with TCB \\ 
                  &  Reordering Statements \\
Allatori  &  Method Reordering \\
          &  Instruction Substitution\\
          &  Reordering Statements\\
          &  Replacing if(non) Null Instructions with TCB\\
          &  Code Clone Type IV \\
          &  Using Opaque Predicate\\
          &  Dead Code Insertion\\
          &  Branch Inversion\\
DashO  &  Opaque Branch Insertion \\
       &  Dead Code Insertion \\
       &  Reordering Statements\\
       &  Goto Instruction Augmentation\\
       & Insert Dummy Loop \\
          &  Replacing if(non) Null Instructions with TCB\\
JMOT  &   Data Sanitization  \\        
JFuscator &  Irrelevant Code Insertion \\
          &  Instruction Substitution\\

\noalign{\smallskip}\hline
\end{tabular}
\label{tab:test3}
\end{minipage}
\hspace{9em}
\begin{minipage}[b]{0.30\linewidth}
\centering
\begin{tabular}{p{1.5cm}p{4.2cm}}
\hline\noalign{\smallskip}
Obfuscator & Obfuscation Technique  \\
\noalign{\smallskip}\hline\noalign{\smallskip}
JBCO &  Reordering the load instructions above the if instructions\\
          &  Replacing if(non) Null Instructions with TCB\\
     &  Adding Dead Code Switch Stmt.\\
     &  Building API Buffer Methods\\
     &  Building Lib Buffer Classes\\
     &  Goto Instruction Augmentation\\
     &  Converting Branches to jsr Instns.\\
     &  Finding and Reusing Duplicate Sequences \\
     &  Disobeying Constructor Conventions \\
     & Partially Trapping Switch Stmts. \\
     & Combining Try Blocks with Catch Blocks \\
     & Indirecting if Instructions \\
Sandmark  &  Boolean Splitter \\
          &  Reordering Expressions \\
          &  Building API Buffer Methods \\
          &  Split Objects \\
           &  Class Splitter  \\
           &  Interleave Methods \\
          &  Inline Methods \\
          &  Convert Reducible to Irreducible Flow Graph \\
          &  Reordering Statements \\
          &  Opaque Branch Insertion \\
          &  Dynamic Inliner \\
          &  Dead Code Insertion \\
\noalign{\smallskip}\hline
\end{tabular}
\label{tab:4}
\end{minipage}
\hfill

\end{table}
\fi

\subsection{JBCO}
JBCO \cite{RefJ51}, \cite{RefJ58} is an obfuscator developed by Sable Group, KTH University, that proposes a set of techniques specifically for bytecode obfuscation. The tool is  a part of the \textit{Soot} framework. We were not able to successfully evaluate all the techniques due to dependency errors while running the tool. However, the techniques they have created have been discussed in detail with examples in \cite{RefJ58}. One thing to note is that depending on the specification of the virtual machine in which bytecode is run, some techniques may  cease to work. 
\begin{enumerate}
\item \textit{Reordering load instruction above if instructions.} This technique moves a local variable common to both the \textit{if} and \textit{else} condition to outside the \textit{if-else} statement, thus removing redundant code along the branches. 

\item \textit{Adding Dead Code Switch Statements.} It inserts a \textit{switch} statement using an opaquely false predicate.

\item \textit{Finding and Reusing Duplicate Sequences.} This technique manipulates the fact that duplicate bytecode sequences occur in various parts of the code. The code sequences can be replaced with a single switched instance, thus altering the control flow as well as reducing the code size. 

\item \textit{Partially Trapping Switch Statements.} Traps are conceptually similar to exception handlers, except that they are specific to bytecode. One way to obfuscate using trap handlers is to encapsulate a sequence of bytecode instructions that may belong to disparate set of code blocks, into a single trap handler. For instance, code blocks belonging to case statements of a switch-case block can be handled by a single trap handler.

\item \textit{Disobeying Constructor Conventions.} This technique manipulates invocations to super class constructor calls. For instance, invocation to a super class constructor is placed inside a try-catch block. 

\item \textit{Combining Try Blocks with Catch Blocks.} While the try block must be followed by the catch block in the source code, at the bytecode, the ordering of the blocks is irrelevant. For the same reason, multiple try-catch blocks maybe combined, or a catch block may precede a try block. 

\item \textit{Indirecting if Instructions.} In this technique, the \textit{if} branch is indirected through a \textit{goto} instruction. To avoid the \textit{goto} from being removed by the compiler, it is enclosed within a try-catch block. 

\item \textit{Building API Buffer Methods.} In this technique, a buffer method is inserted between the caller and the invocation to a Java API call. 

\item \textit{Building Library Buffer Classes.} The obfuscator inserts a buffer class (with methods) in between a caller and the callee library class, thus introducing another level of indirection. 

\item \textit{ Goto Instruction Augmentation.} It splits a method into two sequential parts, uses \textit{goto} to change the control flow of the program. The parts may optionally be reordered.  

\item \textit{Converting branches to jsr instructions.} \textit{jsr} instruction is similar to \textit{goto} the only difference being that jsr pushes a return address to the stack. In this technique,  \textit{if} and \textit{goto} targets are replaced with \textit{jsr} instructions.  The return address is saved in a register for use after returning from a jsr jump. 
 \textit{jsr} is supported only in \textit{Java} version 6.0 and below as per the Oracle Documentation. In our experimental evaluation, we were unable to reproduce this obfuscation technique despite lowering the Java version. 
\end{enumerate}

\subsection{DashO}
DashO extensively uses \textit{try-catch} blocks for altering the control flow of a program. The number of \textit{try-catch} blocks can be varied from 1 to 10. It does not obfuscate compiler-generated classes such as the default constructor. DashO also uses \textit{nop} instructions to increase code size.
\subsection{JMOT}
JMOT supports 4 levels of obfuscation - Light, Normal, Heavy and Insane. The tool has no control flow obfuscation techniques, and only supports \textit{data sanitization} to protect strings used in the program. Strings are passed through a series of function calls named as hax0, hax1 \ldots hax7. However, this in effect alters the control flow of a program. We mention this for completeness here. 

\noindent\begin{minipage}{.45\textwidth}
\begin{lstlisting}[frame=tlrb,label=code:example-22-1,numbers=none]{gotoaug}
public void printResult(int 
       result,boolean isPresent){
if(isPresent==false)
  System.out.println("Element not
                      present");
...
}
\end{lstlisting}
\end{minipage}\hfill
\begin{minipage}{.45\textwidth}
\begin{lstlisting}[frame=tlrb, label=code:example-22-2,numbers=none]{gotoaug}
if (isPresent==false) {
  System.out.println(
  BinarySearch.hax0(
  BinarySearch.hax1(
  ...
  BinarySearch.hax7(
  "Element not present")))))))));
  ...
}
\end{lstlisting}
\end{minipage}

\subsection{JFuscator}
JFuscator is the only obfuscator that gives the developer the choice to obfuscate specific methods or the program as a whole. 

\subsection{Sandmark}
Sandmark is an open source tool developed by University of Arizona that can obfuscate Java programs. Though the tool has been discontinued since JDK version 1.4 (in 2004), we ported the tool to support JDK 1.8. 
To obfuscate, Sandmark provides a list of techniques to choose from. From a total of 15 techniques, we identify 4 that are novel. Given below are the techniques previously seen in literature, but noted below due to minor implementation differences. 

\begin{enumerate}
\item \textit{Interleave Methods.} (Called as \textit{Method Merger in Sandmark}). In the tool, this is implemented as two different techniques - one that interleaves \textit{public static} methods with the same signature and the other that interleaves all methods. 
\item \textit{Inline Methods.} (Called as \textit{Inliner} in Sandmark). In the tool only \textit{static} methods are inlined.
\item \textit{Simple Opaque Predicate.} This is a composition of the techniques \textit{Extending Conditionals} and \textit{Adding Redundant Operands}.
\item \textit{Static Method Bodies.} This is same as \textit{Building API Buffer Methods}, the difference being that it splits all non-static methods into a \textit{static} helper method that will be invoked by a non-static stub.
\item \textit{Branch Inversion.} This technique is a specific implementation of the technique called \textit{Reordering Expressions}, where the conditional expression of an \textit{if} statement is negated.
\item \textit{Buggy Code.} Similar to \textit{Dead Code Insertion}, except that a copy of a basic block is taken to introduce a bug which is then guarded by an opaquely false predicate
\end{enumerate}

The following techniques are novel:
\begin{enumerate}
\item \textit{Dynamic Inliner} (Similar to Inline Methods) This technique uses \textit{instanceof} checks to perform method inlining at runtime.
\item \textit{Boolean Splitter.} It modifies all uses and definitions of \textit{boolean} variables and arrays. In the example below, the boolean \textit{isPresent} is obfuscated.

\begin{minipage}{.43\textwidth}
\begin{lstlisting}[frame=tlrb,label=code:example-14-1, numbers=none]{booleansplitter}
public void printResult(int result,
            boolean isPresent){
  if(isPresent==false) 
  ...
}
\end{lstlisting}
\end{minipage}\hfill
\begin{minipage}{.43\textwidth}
\begin{lstlisting}[frame=tlrb, label=code:example-14-2, numbers=none]{booleansplitter}
public void printResult(int arg1, 
                boolean arg2){
  boolean bool1 = (int)
        (Math.random() * 2.0D); 
  boolean arg1=false;
  if (!(arg1 ^ bool1)) 
  ...
}
\end{lstlisting}
\end{minipage}

\item \textit{Split Objects.} An object of a class is split into two and are linked together by a field. In the example below, the \textit{FindElement} object \textit{ob1} is split. In the obfuscated program, another object called \textit{next0} is created. 

\noindent\begin{minipage}{.43\textwidth}
\begin{lstlisting}[frame=tlrb,basicstyle=\scriptsize,label=code:example-20-1, numbers=none]{splitclasses}
class BinarySearch extends 
                FindElement{
 main(){
  FindElement ob1=new 
             FindElement();
  ...
} }
\end{lstlisting}
\end{minipage}\hfill
\begin{minipage}{.43\textwidth}
\begin{lstlisting}[frame=tlrb,basicstyle=\scriptsize, label=code:example-20-2, numbers=none]{splitclasses}
class FindElement{
  public FindElement0 next0;
  public FindElement(){
    if (this.next0 == null)
      this.next0 = 
            new FindElement0();
}
\end{lstlisting}
\end{minipage}

\item \textit{Class Splitter.} The body of a class is split in such a way that methods and fields of the class are moved into its superclass. In the example, the \textit{printResult} method of \textit{BinarySearch} class is moved to its superclass.

\begin{minipage}{.43\textwidth}
\begin{lstlisting}[frame=tlrb,basicstyle=\scriptsize,label=code:example-32-1, numbers=none]{branchinversion}
class FindElement {
  ...
}
class BinarySearch 
        extends FindElement{
  public void printResult(...){
  ...
} }
\end{lstlisting}
\end{minipage}\hfill
\begin{minipage}{.43\textwidth}
\begin{lstlisting}[frame=tlrb,basicstyle=\scriptsize, label=code:example-32-2, numbers=none]{branchinversion}
class FindElement {
  public void printResult(...){
  ...
} }
class BinarySearch 
        extends FindElement{
}\end{lstlisting}
\end{minipage}

\end{enumerate}

\section{Discussion}

From our study we identified 36 unique techniques from literature and 7 from tools. Three of the most popular and well-supported commercial obfuscators Zelix, Allatori and DashO implement only 13 of the 36 techniques. Thus there appears to be a gap between the theory and practice of CFO. 
\renewcommand{\arraystretch}{1.2}
\begin{table}[htp!]
\caption{Classification \& Analysis Overview}
\label{tab:1}       
\scriptsize
\begin{tabular}{p{.2cm}p{1.6cm}p{4.6cm}p{1.1cm}p{1.1cm}p{1.1cm}p{1.1cm}p{2.4cm}}
\hline\noalign{\smallskip}
 &Component & Obfuscation Technique & Liter. & Tools & Both & DR & Level \\
\noalign{\smallskip}\hline\noalign{\smallskip}
1. & Expression & Extending Conditionals & \cmark & \cmark &\cmark & N &Opaque Predicate\\ 
& &Adding Redundant Operands & \cmark & \cmark &\cmark & N & Opaque Predicate\\
& &Reordering Expressions & \cmark & \cmark & \cmark & N & Ordering\\
2.& Statement & Reordering Statements & \cmark & \cmark & \cmark & N & Ordering\\
& &Remove Lib. and Program Idioms&\cmark & \cmark & \cmark & N & Substitution\\ 
& &Instruction Substitution&\cmark & \cmark & \cmark & N & Substitution \\ 
 & &Replacing \textit{if(Non) Null} Instructions With \textit{Try-Catch}  Block & \cmark & \cmark & \cmark & N & Substitution\\
& &Converting Branches to \textit{jsr} Instr. & \cmark & \cmark & \cmark & N & Substitution\\
3.& Basic Block & Opaque Branch Insertion& \xmark & \cmark & \xmark & N & Opaque Predicate \\
& &Dead Code Insertion & \cmark & \cmark & \cmark & N & Opaque Predicate\\
& &Adding Dead Code Switch Stmts. & \cmark & \cmark & \cmark & N & Opaque Predicate\\
 & &Reordering Loops & \cmark & \xmark & \xmark & N & Ordering \\
 & & Reordering Code Blocks & \cmark & \xmark & \xmark & N & Ordering \\
 & & Finding and Reusing Duplicate Seq. & \cmark & \cmark & \cmark & N & Ordering \\
 & &Reorder Load Instrs. Above \textit{if} Instr.& \cmark & \cmark & \cmark & N & Ordering\\
& &Loop Fission & \cmark & \xmark & \xmark & N & Loop Transf.\\
& &Loop Blocking & \cmark & \xmark & \xmark & N & Loop Transf. \\
& &Loop Unrolling & \cmark & \xmark & \xmark & N & Loop Transf. \\
& &Intersecting Loop & \cmark & \xmark & \xmark & Y &  Loop Transf.\\
& &Replace with Equivalent Codes & \cmark & \cmark & \cmark & N &  Substitution\\
& & Code Clone Type IV & \cmark & \cmark & \cmark & N & Substitution\\
& &Basic Block Fission & \cmark & \xmark & \xmark & Y & Code Insertion\\
& &Insert Dummy Loop & \cmark & \cmark & \cmark & N & Code Insertion \\
& & Goto Instruction Augmentation & \cmark & \cmark & \cmark & Y & Code Insertion\\
& & Irrelevant Code Insertion & \cmark  & \cmark & \cmark & N & Code Insertion \\
& &Control Flow Flattening& \cmark & \xmark & \xmark & N & Code Insertion\\
& &Boolean Splitter&\xmark & \cmark & \xmark & N & Code Insertion\\
& & Convert Reducible to Non-reducible Flowgraph & \cmark & \cmark &\cmark & Y & Code Insertion\\
& & Partially Trapping Switch Stmts & \cmark & \cmark & \cmark & Y &  Code Insertion \\
& & Disobeying Constructor Conventions & \cmark & \cmark & \cmark & Y & Code Insertion \\
& & Combining Try Blocks with Their Catch Blocks & \cmark & \cmark & \cmark & N & Code Insertion \\
& & Indirecting if Instructions & \xmark & \cmark & \xmark & Y & Code Insertion \\
4.& Method & Inline method & \cmark & \cmark & \cmark & N & Method Transf.\\
& & Outline Method & \cmark & \xmark & \xmark & N & Method Transf.\\
& & Clone Method & \cmark & \xmark & \xmark & N & Method Transf. \\
& & Interleave Methods & \cmark & \cmark &\cmark & N & Method Transf.\\
& & Dynamic Inliner& \xmark & \cmark & \xmark & N & Method Transf.\\
& & Building API Buffer Methods& \xmark & \cmark & \xmark & N & Method Transf.\\
& &Table Interpretation  & \cmark & \xmark & \xmark & Y & Method  Transf.\\
& &Parallelizing the Code & \cmark & \xmark & \xmark & N & Method Transf.\\ 

5.& Class  & Split Objects & \xmark & \cmark & \xmark & N & Class Transf.\\
& & Class Splitter & \xmark & \cmark & \xmark & N & Class Transf.\\

& & Building Library Buffer Classes& \cmark  & \cmark & \cmark & N &  Class Transf.\\

\noalign{\smallskip}\hline
\end{tabular}
\end{table}

Table \ref{tab:1} furnishes a classification of the techniques based on the component (or level) of a program at which an obfuscation transformation is applied. We identify 5 different components (or levels) of a program (column 1) - Expression (such as conditional expressions), Statement, Basic Block, Method, Class. We indicate whether the technique is proposed in the literature, tools or both (columns 3-5). In column 6, we note whether these techniques allow a program to be decompiled into a source code equivalent, thus facilitating reverse engineering. Finally, we also note the obfuscating paradigm used in column 7. The foundation for the paradigms is obtained from a prior study by Colberg et al. on taxonomy of obfuscating transformations \cite{RefJ26}.

We assess the potency of these techniques. Colberg et al. \cite{RefJ26} defines potency as a measure of how difficult it is for a human reader to understand the obfuscated program. We augment that measure to include the ability of a decompiler to decompile an obfuscated bytecode into its equivalent source. This is because such techniques will inherently lack structure and thus will be unreadable. We determine that Table Interpretation (Virtualization) is the most potent as the knowledge of how the interpreter will execute the program is with the developer of the obfuscator itself. However, implementing such an obfuscator incurs a huge overhead. In all the other techniques, manual disassembly of the program can leak (at least partially) its proprietary knowledge. 

12 techniques proposed in the literature is not available in any of the tools. 8 of those are transformations applied to a basic block. Basic block transformations are difficult to implement due to the challenge entailed in preserving inter-block dependencies. However, they are also the most potent. Any technique that generates a non-structured control flow at the bytecode level cannot be decompiled. We identify that the following 7 techniques causes a decompiler to fail - \textit{Goto Instruction Augmentation, Conversion of Reducible to Irreducible Flow Graphs, Intersecting Loops, Basic Block Fission, Partially Trapping Switch Statements, Disobeying Constructor Conventions} and \textit{Indirecting if Instructions}. \textit{Loop blocking}, though it can be decompiled, is potent, as it makes use of nested loops. Nested loops make use of nested \textit{labels} and \textit{gotos} that deters the readability of a program for a human analyst.

All the other techniques that use ordering, substitution, class and method transformation paradigms are effective to create variants of a program, but does not deter the readability of a program. Literature cites method reordering as a CFO technique. However, method reordering only alters the class structurally and does not alter its control flow. Hence this cannot be classified as a CFO technique. We however list it here for completeness. Transformations on expressions and statements are the most common and easiest to implement. \textit{Opaque predicates} can be potent if the predicate used to secure a conditional is potent.

None of the techniques can be deobfuscated without the original program. However, techniques that use opaque predicates such as dead code, opaque branch insertion, redundant operand etc. are common compiler optimizations that can be used to deobfuscate. A deobfuscator can also remove control flow edges that generate non-reducible flow graphs, to generate structured programs so as to enable a decompiler to generate source code.

Of the 6 tools we evaluated, Zelix, Allatori, DashO and JMOT can be used by software developers to control flow obfuscate their code. We determine that Zelix is the most potent as it is the only tool that implements language breaking transformations. Though Allatori and DashO are comparable in their implementations, programs generated by Allatori mask their obfuscation techniques well. DashO relies on try-catch blocks (TCB). They insert a varying number of TCBs depending on the level of obfuscation. This approach does not substantially mar the readability of a program. Although Sandmark provides the most number of techniques, it is not a practically viable tool for commercial purposes. This is because, the tool is no longer supported and not tested for versions of Java greater than 1.4. Each tool has a unique way of inserting dead code, which is useful to identify the obfuscator. Though JBCO has a good lineup of techniques, it has a broken code base at the time of writing this paper and hence not usable. JMOT is a purely command-line tool, does not support control flow obfuscation

The following techniques are common to most tools - \textit{Extending Conditionals, Adding Redundant Operands, Opaque Predicate Insertion, Dead Code Insertion, Irrelevant Code Insertion, Reordering Expressions, Reordering Statements, Replacing \textit{if}(non)null instructions with \textit{try-catch} blocks, and Goto Augmentation}. Aside from the 8 techniques that are decompiler resistant, none of the other techniques provide enough security to prevent reverse engineering attacks. However, majority of the tools implement none of those techniques, allowing an adversary to defeat obfuscation.

\section{Conclusion}\label{sec:Conclusion}
In this research, we have surveyed existing literature and evaluated 13 tools to identify the different control flow obfuscation techniques and their implementations. We have identified 36 unique techniques from literature and 7 from tools. The three most popular commercial tools use only 13 of the 36 techniques, thus showing a lag between the theory and practice of control flow of obfuscation. We classify the 43 techniques into 5 types based on the component of a program on which obfuscation is applied. We also discuss the obfuscating paradigm used. Based on our analysis, we determine that transformations on a basic block are the most potent. We identify 8 techniques that renders a program unreadable both to a human analyst and a program decompiler. 7 of those techniques use language breaking transformations to make a program unstructured. All the other obfuscations are effective to create semantic variants and does not deter readability of a program. Most of the obfuscators do not implement potent techniques defeating the reliability of obfuscation. Each tool inserts their own proprietary dead code, which a deobfuscator can use as a feature to detect the tool. We identify the most commonly used techniques in tools and also flag those that are not implemented. When using a composition of obfuscations, there is a drastic increase in code size and complexity. Hence determining the optimal amount of obfuscation such that the  run-time performance and code complexity is also optimal while preserving potency is an area that requires further work.


\end{document}